\documentclass[sigconf]{acmart}

\AtBeginDocument{%
  \providecommand\BibTeX{{%
    \normalfont B\kern-0.5em{\scshape i\kern-0.25em b}\kern-0.8em\TeX}}}

\usepackage[inline]{enumitem}
\makeatletter
\newcommand{\sol}[1]{%
   \protected@write \@auxout {}{\string \newlabel {sol:#1}{{S#1.}{\thepage}{#1}{sol:#1}{}} }%
   \hypertarget{sol:#1}{(S#1)}%
}\makeatother

\copyrightyear{2023}
\acmYear{2023}
\setcopyright{rightsretained}
\acmConference[CoNEXT-SW '23]{Proceedings of the CoNEXT Student Workshop 2023}{December 8, 2023}{Paris, France}
\acmBooktitle{Proceedings of the CoNEXT Student Workshop 2023 (CoNEXT-SW '23), December 8, 2023, Paris, France}
\acmDOI{10.1145/3630202.3630231}
\acmISBN{979-8-4007-0452-9/23/12}

\begin{document}

\title{Using SRv6 to access Edge Applications in 5G Networks}

\author{Louis Royer}
\orcid{0009-0008-4492-2804}
\affiliation{%
  \institution{IRIT, Université de Toulouse}
  \city{Toulouse}
  \country{France}
}
\email{louis.royer@irit.fr}

\author{Emmanuel Lavinal}
\orcid{0000-0002-5899-0444}
\affiliation{%
  \institution{IRIT, Université de Toulouse}
  \city{Toulouse}
  \country{France}
}
\email{emmanuel.lavinal@irit.fr}

\author{Emmanuel Chaput}
\orcid{0000-0001-7100-4867}
\affiliation{%
  \institution{IRIT, Université de Toulouse}
  \city{Toulouse}
  \country{France}
}
\email{emmanuel.chaput@irit.fr}

\begin{abstract}

With the emergence of Multi-Access Edge Computing in 5G and beyond, it has become essential for operators to optimize the data path for the end-user while ensuring resources are used according to their policy. In this paper, we review existing solutions to access edge resources, underline their limits, and propose the use of Segment Routing over IPv6~(SRv6) in a 5G/edge architecture. 
\end{abstract}



\maketitle

\section{Introduction}

Edge computing brings computing resources closer to the users and devices, resulting in lower latency, improved reliability, and support for bandwidth-intensive applications. This technology is particularly relevant in the context of 5G and beyond mobile networks as it increases quality of experience for end-users and utilizes more efficiently the mobile backhaul and core networks \cite{MEC-Survey-17}. 
Although some mobile network operators have initiated pilot projects and deployments of edge computing solutions, these deployments are often partial or for specific use cases, such as industrial IoT or content delivery.

\begin{figure}[ht]
  \centering
  \includegraphics[width=\linewidth]{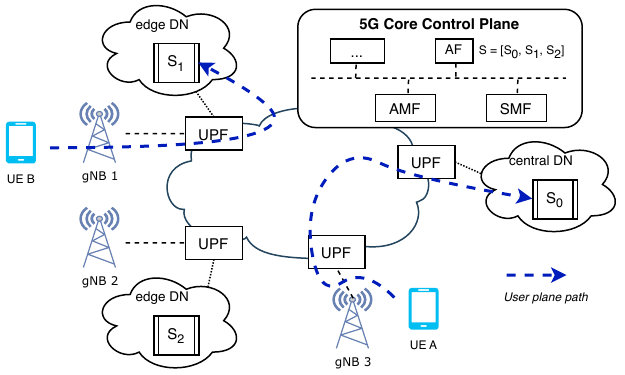}
  \caption{Edge computing in a 5G context.}
  \Description{A 5G Network with 3 instances (S0, S1, S2) of a service (S) and 2 User Equipments (UE A, UE B). UE A is accessing S0 on central Data Network using gNB 3, UE B is accessing S1 on an Edge Data Network which is located near gNB1 that UE B is using.}
  \label{fig:context-5g-edge}
\end{figure}

As illustrated in Figure~\ref{fig:context-5g-edge}, we consider 5G and beyond networks in which applications have multiple instances that are deployed on different edge data networks.
The Mobile Network Operator (MNO) has to provide access to the proper instance at the right location and at the right time, according to a specific policy (e.g. to respect an SLA).
In this context, many issues have to be addressed such as \textit{instances management} (to create, destroy, or migrate application instances), \textit{binding} (to map the service requested by a User Equipment (UE) to a specific instance ID that will provide the service), and \textit{access} (to ensure the data plane is configured such that the UE has access to the bound instance).
In our current work, we mainly focus on the dynamic access to instances, arguing that a solution should meet five central requirements:
\begin{enumerate}[label={(R\arabic*)}, ref={R\arabic*.}]
\item \label{req:1} \textit{5G Control Plane}: the solution has to be compatible with existing 5G Control Plane.
\item \label{req:2} \textit{Operator control}: the MNO should have control on the binding and on the effective access. This allows the operator to enforce QoS according to the SLA defined for the slice, and to manage efficiently resources.
\item \label{req:3} \textit{User mobility}: users should be able to geographically move, with, as a possible consequence, binding and data path updates.
\item \label{req:4} \textit{Runtime update}: new services and new instances can be created dynamically. When this is appropriate, bindings should be updated, resulting in a seamless use of this instance.
\item \label{req:5} \textit{Scalability}: the approach should scale with the number of mobile users and the number of instances, despite the granularity of QoS rules.
\end{enumerate}

\section{Related Work}

\begin{table*}
\caption{Existing solutions to access edge application instances}
\label{table:existing-solutions}
\small{
\begin{tabular}{llllll}
    \toprule
        Solution &
        \ref{req:1} \textit{5G CP} &
        \ref{req:2} \textit{Operator control} &
        \ref{req:3} \textit{Runtime update} &
        \ref{req:4} \textit{User Mobility} &
        \ref{req:5} \textit{Scalability}  \\
    \midrule
    \ref{sol:0} \textit{Hardcoded}   & Independent & No                  & No                                   & No      & No          \\
    \ref{sol:1} \textit{App. based}  & Independent & Developer dependant & Developer dependant                  & Partial & No          \\
    \ref{sol:2} \textit{DNS based}   & Compatible  & Developer dependant & Partial                              & Partial & Partial     \\
    \ref{sol:3} \textit{ULCL}        & Integrated  & Yes                 & Complex UPF/SMF rules                & Yes     & No          \\
    \ref{sol:4} \textit{Multihoming} & Integrated  & Partial             & PDU Session reconf. (UE IPv6 Prefix) & Yes     & Partial     \\
    \bottomrule
\end{tabular}
}
\end{table*}

The naive solution to connect a UE with an instance is to hardcode the instance ID into the client application \sol{0}. This solution is the least flexible. It can make sense when there is a unique instance for the service, but cannot fit most usecases.

In their white paper, ETSI has suggested two ways for the UE to connect with an instance \cite{ETSI-WP20ed2_MEC}:
\sol{1}~the application developer has the responsibility for making the instance ID available to the UE (App. based solution); \sol{2}~the UE discovers the instance ID (DNS based solution).
With application based solutions (e.g. URL redirection), binding updates depend entirely on the developer. Efforts can be made to take user mobility into account, but this solution relies on a central instance used for initial access and therefore the scalability criteria is not met.
Regarding DNS based approaches, they can scale with the number of users, but they are not adapted to user mobility due to cache invalidation delay.

In 5G networks, an alternative solution is to delegate the binding responsibility to the network operator and to use traffic steering to reach the target instance \cite{ETSI-WP28ed1_MEC}. Solutions based on traffic steering avoid the need for the UE to know the precise identifier of the instance, offering additional flexibility for resource management and user mobility.
An Uplink~Classifier~\sol{3} can be inserted in the data path by the SMF to perform traffic steering. This solution works with a limited number of mobile users and service instances, but the complexity of the steering rules and the number of control messages required for runtime updates constitute an important drawback.
Alternatively, IPv6 multihoming~\sol{4} can be used to route PDUs according to the source IP prefix, but this information may not be sufficient to select the right instance at the right time.

Table~\ref{table:existing-solutions} summarizes existing solutions to access application instances with respect to the requirements exposed in the previous section.

\section{Proposal and preliminary experiments}

\begin{figure}[ht]
  \centering
  \includegraphics[width=\linewidth]{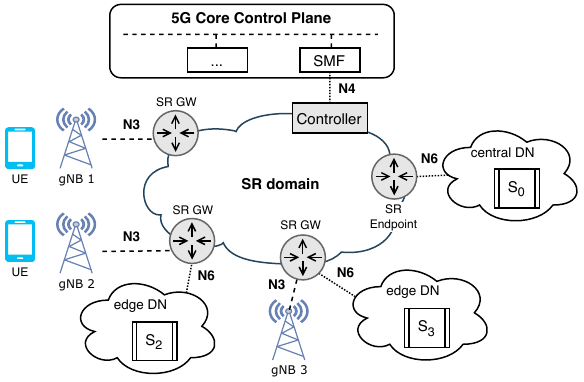}
  \caption{Integration of SRv6 within 5G Networks.}
  \Description{A 5G Network where UPFs have been replaced by SR Gateways and a Controller.}
    \label{fig:5g-edge-sr-proposal}
\end{figure}

In order to support dynamic access and updates to edge application instances, we propose to rely on SRv6 for the Mobile User Plane \cite{rfc9433}.
In this approach, SRv6 is used as the user plane of mobile networks, allowing operators to explicitly indicate a route for the packets to and from the mobile node.
Moreover, the data paths can be controlled with a high granularity while having lightful processing of packets at the core of the network.
Indeed, SRv6 integrates both the application data path and the underlying transport layer into a single protocol, meaning that all traffic steering operations can be performed by any of the SR-aware intermediate nodes in a stateless fashion.
The general architecture we adopt is depicted in Figure~\ref{fig:5g-edge-sr-proposal} in which SR gateways map inbound and outbound GTP-U traffic into SRv6 (i.e. the ``Enhanced Mode with Unchanged gNB GTP-U Behavior'' in \cite{rfc9433}).
We have also added a controller to the SR domain in order to dynamically update rules on the SRv6 source nodes that enforce the data paths following the operator's policy.
Our controller is seen by the SMF as a single UPF, thus greatly simplifying
the states maintained by the SMF and avoiding complex interactions with many UPFs to setup, chain and tear down GTP-U tunnels.
Processing of packets is also simplified for backbone routers since the data paths are set only on edge nodes, thus avoiding state management, and reducing the amount of control messages required for runtime updates.
In addition, this solution allows function chaining for no additional cost.

We have developed a testbed using UERANSIM \cite{UERANSIM} for the Access Network, free5GC \cite{free5GC} for the Control Plane of the Core Network, and a Segment Routing domain composed of a controller and multiple Linux based SRv6-enabled routers.
The controller is implemented in Go and includes a Packet Forwarding Control Protocol (PFCP) interface to communication with free5GC's SMF.
For the moment, rules (including GTP-U/SRv6 mapping rules) are statically defined in a configuration file.
We have successfully executed on this testbed a usecase in which a UE accesses two different application instances depending on the 5G slice it connects to.

\section{Future work}
Future work intend to enhance our current implementation to allow the controller to dynamically reconfigure the SR~gateways, so that it can handle runtime updates. Moreover, we are working on user mobility to be able to integrate 5G~handover procedures within our controller.

\bibliographystyle{ACM-Reference-Format}
\balance
\bibliography{biblio}


\begin{thebibliography}{6}


\ifx \showCODEN    \undefined \def \showCODEN     #1{\unskip}     \fi
\ifx \showDOI      \undefined \def \showDOI       #1{#1}\fi
\ifx \showISBNx    \undefined \def \showISBNx     #1{\unskip}     \fi
\ifx \showISBNxiii \undefined \def \showISBNxiii  #1{\unskip}     \fi
\ifx \showISSN     \undefined \def \showISSN      #1{\unskip}     \fi
\ifx \showLCCN     \undefined \def \showLCCN      #1{\unskip}     \fi
\ifx \shownote     \undefined \def \shownote      #1{#1}          \fi
\ifx \showarticletitle \undefined \def \showarticletitle #1{#1}   \fi
\ifx \showURL      \undefined \def \showURL       {\relax}        \fi
\providecommand\bibfield[2]{#2}
\providecommand\bibinfo[2]{#2}
\providecommand\natexlab[1]{#1}
\providecommand\showeprint[2][]{arXiv:#2}

\bibitem[{free5GC Project}(2023)]%
        {free5GC}
\bibfield{author}{\bibinfo{person}{{free5GC Project}}.}
  \bibinfo{year}{2023}\natexlab{}.
\newblock \bibinfo{title}{{free5GC}}.
\newblock
\newblock
\urldef\tempurl%
\url{https://www.free5gc.org/}
\showURL{%
\tempurl}


\bibitem[Güngör et~al\mbox{.}(2023)]%
        {UERANSIM}
\bibfield{author}{\bibinfo{person}{Ali Güngör} {et~al\mbox{.}}}
  \bibinfo{year}{2023}\natexlab{}.
\newblock \bibinfo{title}{{UERANSIM}}.
\newblock
\newblock
\urldef\tempurl%
\url{{https://github.com/aligungr/UERANSIM}}
\showURL{%
\tempurl}


\bibitem[Kekki et~al\mbox{.}(2018)]%
        {ETSI-WP28ed1_MEC}
\bibfield{author}{\bibinfo{person}{Sami Kekki} {et~al\mbox{.}}}
  \bibinfo{year}{2018}\natexlab{}.
\newblock \bibinfo{booktitle}{\emph{MEC in 5G networks}}.
\newblock Number~28 in \bibinfo{series}{ETSI White Papers}.
  \bibinfo{publisher}{MEC ISG}, \bibinfo{address}{ETSI}.
\newblock
\showISBNx{979-10-92620-22-1}
\newblock
\shownote{1\textsuperscript{rst} ed}.


\bibitem[Matsushima et~al\mbox{.}(2023)]%
        {rfc9433}
\bibfield{author}{\bibinfo{person}{Satoru Matsushima},
  \bibinfo{person}{Clarence Filsfils}, \bibinfo{person}{Miya Kohno},
  \bibinfo{person}{Pablo Camarillo}, {and} \bibinfo{person}{Daniel Voyer}.}
  \bibinfo{year}{2023}\natexlab{}.
\newblock \bibinfo{title}{{Segment Routing over IPv6 for the Mobile User
  Plane}}.
\newblock \bibinfo{howpublished}{RFC 9433}.
\newblock
\urldef\tempurl%
\url{https://doi.org/10.17487/RFC9433}
\showDOI{\tempurl}


\bibitem[Sabella et~al\mbox{.}(2019)]%
        {ETSI-WP20ed2_MEC}
\bibfield{author}{\bibinfo{person}{Dario Sabella} {et~al\mbox{.}}}
  \bibinfo{year}{2019}\natexlab{}.
\newblock \bibinfo{booktitle}{\emph{Developing Software for Multi-Access Edge
  Computing}}.
\newblock Number~20 in \bibinfo{series}{ETSI White Papers}.
  \bibinfo{publisher}{MEC ISG}, \bibinfo{address}{ETSI}.
\newblock
\showISBNx{979-10-92620-29-0}
\newblock
\shownote{2\textsuperscript{nd} ed}.


\bibitem[Taleb et~al\mbox{.}(2017)]%
        {MEC-Survey-17}
\bibfield{author}{\bibinfo{person}{T. Taleb}, \bibinfo{person}{K. Samdanis},
  \bibinfo{person}{B. Mada}, \bibinfo{person}{H. Flinck}, \bibinfo{person}{S.
  Dutta}, {and} \bibinfo{person}{D. Sabella}.} \bibinfo{year}{2017}\natexlab{}.
\newblock \showarticletitle{On Multi-Access Edge Computing: A Survey of the
  Emerging 5G Network Edge Cloud Architecture and Orchestration}.
\newblock \bibinfo{journal}{\emph{IEEE Communications Surveys \& Tutorials}}
  \bibinfo{volume}{19}, \bibinfo{number}{3} (\bibinfo{year}{2017}),
  \bibinfo{pages}{1657--1681}.
\newblock
\urldef\tempurl%
\url{https://doi.org/10.1109/COMST.2017.2705720}
\showDOI{\tempurl}


\end{thebibliography}
\end{document}